\documentclass[11pt,a4paper]{scrartcl}
\usepackage{amsmath,amssymb}
\usepackage[T1]{fontenc}
\usepackage{lmodern}
\usepackage{braket}
\usepackage{comment}
\usepackage{graphicx}
\newcommand{\Tr}{\mathrm{Tr}}

\title{Classical solution for
ghost D-branes in string field theory
}

\author{Syoji Zeze\footnote{ztaro21@gmail.com}\\
	Yokote Seiryo Gakuin High School\\
	147-1 Maeda, Osawa, Yokote, 013-0041 Japan}
\date{}

\begin{document}

\maketitle
\begin{abstract}
A ghost D-brane has been proposed as an object that cancels the effects of a D-branes. 
We construct a classical solution with an arbitrary number of D-branes
and ghost D-branes in the context of open string field theory. 
Cancellation of BRST cohomolgy between D-branes and ghost 
D-branes is shown.   
\end{abstract}

\section{Introduction}

Recent developments in open string field theory (OSFT)~\cite{Witten:1985cc} 
have proved that it is able to describe wide variety of open string backgrounds.  
A prominent example is the solution found by Erler and Maccaferri~\cite{Erler:2014eqa} 
which covers wide range of open string backgrounds such as
marginal deformation, D-brane lump and multiple D-branes.  
More recently, 
Kojita, Maccaferri, Masuda and Schnabl have incorporated 
topological defects 
between boundary conformal field theories (BCFTs)~\cite{Cardy:2004hm} 
into OSFT context~\cite{Kojita:2016jwe}.    
A common belief behind those developments is that OSFT 
governs the landscape of (hopefully all possible) BCFTs. 
Studies on explicit background are still important to 
understand such landscape.  

In this paper, we are interested in \textit{ghost D-branes}, which are
rather different from D-branes in conventional BCFTs.   
Originally, ghost D-branes (gD-branes) were introduced by Okuda and Takayanagi
as objects that 
cancel the effect of D-branes~\cite{Okuda:2006fb} and studied subsequently by 
some authors~\cite{Terashima:2006qm,Evans:2006eq}
\footnote{
The author became aware of works~\cite{Vafa:2001qf,Parkhomenko:2001ki,Parkhomenko:2003gy,Parkhomenko:2004ab,Parkhomenko:2008dt,Dijkgraaf:2016lym}
which also deal with branes with negative tension. 
We thank S. E. Parkhomenko for correspondence. }. 
It is characterized by a boundary state with 
opposite sign:
\begin{equation}
\ket{B (gD)}  =  - \ket{B (D)}.
\end{equation}
Then, the amplitude for closed string propagation between D and gD branes is
given by $\bra{B(gD)} \Delta \ket{B(D)}= - \bra{B(D)} \Delta \ket{B(D)}$, where
 $\Delta$ is an inverse of the closed string propagator. Obviously, this has a sign
opposite to the amplitude with two D-branes. 
 In open string channel, these
amplitudes are interpreted as one-loop partition functions of open strings stretched
between branes.  The negative sign
of D-gD partition function can be attributed 
to fermionic path integral. Thus, field 
content on a coincident D-gD pair is 
\begin{equation}
	\begin{pmatrix}
	\varphi_1 & \psi\\
	\psi^{\dagger} &  \varphi_2
	\end{pmatrix}
	\label{matrix}
\end{equation}
where $\varphi_1$ and $\varphi_2$ are bosonic while $\psi$
and $\psi^{\dagger}$ are fermionic.    Extension to 
$N$ D-branes and $M$ gD-branes give rise to $U(N|M)$ or $OSp(N|M)$ matrix with similar structure
as \eqref{matrix}. 
The authors of \cite{Okuda:2006fb} showed
that D and gD branes cancel each other in the 
partition function of supermatrix model defined 
by \eqref{matrix}. 
With that result, they claimed that a gD-brane cancels 
D-brane completely therefore a D-gD pair is equivalent to the tachyon (or closed) string vacuum. 
This means that all physical observables cancel between D and gD branes. 
 
Subsequently, several authors encountered gD-branes 
in the efforts of constructing solution for multiple D-branes in 
OSFT~\cite{Murata:2011ep,Hata:2011ke,
	Hata:2012cy,Masuda:2012kt,Erler:2012dz,Erler:2012qn}
\footnote{Refs.~\cite{Baba:2006rs,Baba:2007tc} 
	also rediscovered gD-branes as boundary states.}.
In this context, a gD-brane is given by a classical solution with
negative tension.  It has not been drawn much attentions because of
its peculiar nature.  However, it is worthwhile to mention that single gD-brane was found to be regular solution
in all known literature.  
In particular, the gD-brane solution discovered by Masuda, Noumi
and Takahashi (MNT)~\cite{Masuda:2012kt} passed thorough most stringent
consistency checks ever known.  
On the other hand, a solution for multiple D-branes in universal space, 
which was main 
 aim of the authors of~\cite{Murata:2011ep,Hata:2011ke,
	Hata:2012cy,Masuda:2012kt,Erler:2012dz,Erler:2012qn},
turned out to fail such consistency checks unless quite 
subtle regularization (for example, phantom term) is assumed. 
Therefore, it is worth to investigate gD-branes seriously
in the context of OSFT.

In this paper, we will ask two questions about nature of gD-brane.  
First, we will ask whether it is \textit{physical degrees of freedom.}
Its negative tension implies inconsistency of
bosonic OSFT. Therefore it is important to ask whether there exists a logic to 
exclude gD-brane from the spectrum.   Unfortunately, we will not have conclusive 
answer to this 
question.  We will discuss about possible solutions of 
the problem in terms of gauge fixing.

Second, we will ask \textit{how gD-branes in OSFT differ from 
	the original picture of Okuda and Takayanagi.} 
A crucial difference between them is seen in a way
to cancel D and gD branes.  
In the original picture, the cancellation completely holds at
quantum level since partition function becomes trivial for 
D-gD pairs.
One the other hand, the cancellation in OSFT was confirmed only for
physical observables.   To close the gap between two, we will 
propose another criterion for the cancellation.
Namely, it is cancellation of \textit{BRST cohomology}. 
It will be shown that the BRST cohomology of D-gD pairs vanish
with suitable choice of the string field.    

Before going into details, we briefly sketch main results of our paper.  
MNT gD-brane~\cite{Masuda:2012kt} is  
simply given by a sum of two ``tachyons'':
\begin{equation}
 \Psi_{MNT} = \Psi_{F} + \Psi_{H},
\end{equation}
where $\Psi_{F}$ and $\Psi_{H}$ are Okawa-type analytic 
solutions~\cite{Okawa:2006vm} that are orthogonal with each other:
\begin{equation}
 \Psi_{F} \Psi_{H} = \Psi_{H} \Psi_{F}=0.
\end{equation}
In order to prove D-gD cancellation of cohomology,  we fix $\Psi_{F}$ to be tachyon and 
switch to the vacuum defined by
\begin{equation}
 Q_{F} = Q_{B} + \{\Psi_{F}, *\}.
\end{equation}
Then, a solution for a D-gD pair is identified as
\begin{equation}
 \Psi_{D+gD}  = - \Psi_{F} + \Psi_{H}.
\end{equation}
BRST cohomology around this solution vanishes as we will see, since 
\begin{equation}
	\begin{split}
	Q_{D+gD} & = Q_{F} + \{\Psi_{D+gD}, *\}  \\
	& = Q_{B} + \{\Psi_{H}, *\} 
	\end{split}
\end{equation}
just corresponds to a kinetic operator for ``another'' tachyon.   
Thus, we have shown that a D-gD pair $\Psi_{D+gD}$ has no cohomology therefore 
physical excitations cancel out.  We also 
construct a solution with an arbitrary numbers of D and gD branes:
\begin{equation}
 \Psi_{N,M}  = - \sum_{a=1}^{N} \Sigma_{k_{a}} \Psi_{F} \overline{\Sigma}_{k_{a}}
    +  \sum_{a=1}^{M} \Sigma_{l_{a}} \Psi_{H} \overline{\Sigma}_{l_{a}},
\end{equation} 
where $\Sigma_{k_a}$ and $\overline{\Sigma}_{k_a}$ are the modified BCC 
projectors introduced in~\cite{Erler:2014eqa}. 
It will be shown that
the cohomology not always cancels 
but cancels in a subset of whole string fields.  

This paper is organized as follows. 
In section \ref{sec:MNT}, after a brief review of
the MNT solution~\cite{Masuda:2012kt},  
we will introduce a set of solutions 
which describes the D-gD system.  
Cancellation between D and gD branes will be confirmed
for gauge invariant observables and BRST cohomology.    
In section \ref{sec:multiple}, we will extend this result to
multiple branes.   We will conclude in section \ref{sec:conclusions} with
some discussions.

\section{MNT ghost D-brane} \label{sec:MNT}

\subsection{Solutions}
We begin with a study of the ghost D-brane solution discovered by 
Masuda, Noumi and Takahashi (MNT)~\cite{Masuda:2012kt}. 
As explained in introduction,  it is simply a sum of formal solutions~\cite{Okawa:2006vm} 
of equation of motion around the perturbative vacuum:
\begin{equation}
\Phi_{MNT} = \Psi_{F} + \Psi_{H}, \label{DgDsum}
\end{equation}
where
\begin{equation}
 \Psi_{F} = F c \frac{K}{1-F^2} B c F, \quad \Psi_{H} = H c \frac{K}{1-H^2} B c H, \label{okawa}
\end{equation}
where $c$ and $B$ are elements of the $ K B c $ algebra, and $F$ and $H$ are functions of $K$.  
Since the equation of motion is quadratic, a sum of two Okawa solutions 
never becomes a solution unless their product vanish, i.e.,
\begin{equation}
  \Psi_{F} \Psi_{H} = \Psi_{H} \Psi_{F} = 0.
\end{equation}
We will refer this relation as \textit{orthogonality}.  
Two solutions \eqref{okawa}  
becomes orthogonal when $F H =1$ holds.  
It is also required that both $\Psi_{F}$ and $\Psi_{H}= \Psi_{F^{-1}}$ 
derive the expected values of the classical action (in this paper, normalized to be $-1$)
and vanishing cohomologies.  In addition, the authors
impose a consistency condition 
for the boundary state~\cite{Kiermaier:2008qu} to
the solutions.  They showed an explicit choice
\begin{equation}
 F = \sqrt{\frac{1-p K}{1-q K}}, \quad 
 H = \sqrt{\frac{1-q K}{1-p K}}, \label{FandH}
\end{equation}
where $p$ and $q$ are positive numbers, fulfills
all requirements.   With this choice, they
showed that the solution \eqref{DgDsum} should be identified with a gD-brane. 
In order to confirm this, 
it is convenient to switch to the tachyon vacuum. 
We fix $\Psi_F$ to be the ``reference'' tachyon vacuum\footnote{
As is clear from \eqref{FandH}, there is no essential difference between
$F$ and $H$ since they exchange under $p \leftrightarrow q$. 
$\Psi_{F}$ and $\Psi_{H}$ are gauge equivalent since they derive same
physical observables therefore belong to same class of analytic solutions~\cite{Masuda:2012kt}. 
}
so that the theory is described by an action
\begin{equation}
 S_{F} [\Psi] =\mathrm{Tr}\left[
 \frac{1}{2} \Psi Q_{F} \Psi + \frac{1}{3} \Psi^3
 \right] \label{faction}
\end{equation}
where $Q_{F} = Q_{B} + \{\Psi_{F},*\}$.  Then, equation of motion of 
the action \eqref{faction} is $Q_{F} \Psi + \Psi^2=0$ and we find 
four solutions composed by $\Psi_{F}$ and $\Psi_{H}$:
\begin{align}
 \Psi_{D} & = -\Psi_{F}, \label{Dsolution}\\
 \Psi_{TV}  & = 0, \\ 
 \Psi_{D+gD} & = -\Psi_{F} + \Psi_{H} \label{D+gDsolution}\\
 \quad \Psi{gD} &=  \Psi_{H}. \label{gDsolution}
\end{align}
It is straightforward to confirm that all of them satisfy the equation of motion.  
We identify $\Psi_{D}$, $\Psi_{TV}$, $\Psi_{D+gD}$ and $\Psi_{gD}$ 
as D-brane, tachyon vacuum, 
D-gD pair and gD-brane respectively. 
An evidence of this identification comes from values of tension;
$1$ for D-brane, $0$ for the tachyon vacumm and D-gD pair and 
$-1$ for gD-brane.

\subsection{Gauge invariant observables} 
In this section, we evaluate
three gauge invariant observables: classical action, gauge invariant overlap and boundary state. 
Let $\mathcal{O} (\Psi)$ be a gauge invariant observables. 
Then the cancellation is represented by
\begin{equation}
 \mathcal{O} (\Psi_{D+gD}) = 0,
\end{equation}
where $\mathcal{O}$ is an observable of interest.

The cancellation of the tension is easily established as 
follows:
\begin{equation}
\begin{split}
S_{F} [\Psi_{D+gD}] & = S_{F} [\Psi_{D}] + 
S_{F} [\Psi_{gD}] \\
& = 1 + (- 1 ) \\
& = 0,
\end{split}
\end{equation}
where we have used orthogonality in the first line.

We also confirm cancellation of gauge invariant overlap for closed string~\cite{Ellwood:2008jh,Kawano:2008ry}.  
The evaluation goes straightforwardly as
\begin{equation}
\begin{split}
\bra{I} \mathcal{V}(i)\ket{ \Psi_{D+gD} } & =   
\bra{I} \mathcal{V}(i)\ket{ \Psi_{D} } +
\bra{I} \mathcal{V}(i)\ket{ \Psi_{gD} }  \\
& =   - \bra{I} \mathcal{V}(i)\ket{ \Psi_{F} } +
\bra{I} \mathcal{V}(i) \ket{ \Psi_{H} }    \\
& = 0, 
\end{split}
\label{overlap}
\end{equation}
where $\mathcal{V}$ is a closed string vertex operator and $\bra{I}$ is the identity string field.   In third line, we used the fact that the values of the overlap for are common for tachyons.
Above result \eqref{overlap} can be further extended to cancellation of boundary state
with a help of Ellwood conjecture~\cite{Ellwood:2008jh}:
\begin{equation}
 \bra{\mathcal{V} } c_{0}^{-} \ket{B (\Psi)}-
 \bra{\mathcal{V} } c_{0}^{-} \ket{B (\Psi_0 )}
  =  \bra{I}\mathcal{V}(i)
 \ket{\Psi}, \label{ellwood}
\end{equation}
where $\Psi_{0}$ refers to a classical solution for the reference 
BCFT\footnote{We omit $4 \pi i$ factor in right hand side of \eqref{ellwood}.}.
Similarly, $\Psi$ refers to an open string field in the reference OSFT at the 
perturvative vacuum.  
Applying \eqref{overlap} to above equation we have
\begin{equation}
\begin{split}
\bra{\mathcal{V} } c_{0}^{-} \ket{B ( \Psi_{0} +\Psi_{D+gD})} 
- \bra{\mathcal{V} } c_{0}^{-} \ket{B (\Psi_0 )}
 & =  \bra{I} \mathcal{V}(i)\ket{ \Psi_{D+gD} + \Psi_0 }  \\
 & = 0,
\end{split}
 \label{boundary2}
\end{equation}
where we have set $\Psi_0$ as zero. 
We regard boundary states in left hand side of above equation as
the gauge invariant boundary state introduced by
Kudrna,  Maccaferri and Schnabl with assuming the
uplift to the auxiliary CFT~\cite{Kudrna:2012re}.  
Since their boundary states is linear with respect to open string fields, \eqref{boundary2}
is reduced to
\begin{equation}
\bra{\mathcal{V} } c_{0}^{-} \ket{B (\Psi_{D+gD})} 
=0. 
\end{equation}
As explained in~\cite{Kudrna:2012re}, 
Ellwood conjecture~\eqref{ellwood} essentially contains all 
information of closed string state therefore can be extended to arbitrary closed string state. 
Therefore we have
\begin{equation}
 \ket{B  (\Psi_{D+ gD})} = 0,
\end{equation}
which can be interpreted as cancellation of 
boundary state\footnote{MNT confirmed the cancellation in 
boundary state in rather different way. 
They used KOZ boundary state~\cite{Kiermaier:2008qu} which is neither linear 
nor gauge invariant. 
They showed directory $\ket{B(\Psi_{D})}= \ket{B}$ and $\ket{B(\Psi_{gD})}= - \ket{B}$.   
}.  

\subsection{Cohomology} \label{sec:Hcancel}
As explained in introduction, it is easy to find empty
cohomology of a D-gD pair. Namely, we evaluate the 
kinetic operator as
\begin{equation}
\begin{split}
Q_{D+gD} & = Q_{F} + \{\Psi_{D+gD}, *\} \\
& = Q_{B} + \{\Psi_{F}, *\}+ \{-\Psi_{F} + \Psi_{H}, *\} \\
& = Q_{B} + \{\Psi_{H}, *\}  \\ 
& = Q_{H}. 
\end{split}
\label{DgDcharge}
\end{equation}
This is nothing but a BRST charge shifted by 
 ``another'' tachyon vacuum specified $\Psi_{H}$ therefore has a nontrivial 
 homotopy operator~\cite{Ellwood:2006ba}
\begin{equation}
 A = \frac{1-H^2}{K} B,
\end{equation}
and of course, its cohomology vanishes. 
This means that there are no physical excitations around a D-gD pair
even at quantum level. An OSFT around a D-gD pair is completely 
equivalent to
that around the tachyon vacuum since their cohomlogies are identical.
Thus, our result validate the statement
\textit{A system with a pair of D-brane and ghost D-brane located at the same location
is physically equivalent to the closed string vacuum}~\cite{Okuda:2006fb}
in the context of OSFT.

\subsection{Projections} \label{sec:projections}

Our construction of D-gD system is largely owing to the orthogonality between $\Psi_{F}$ 
and $\Psi_{H}$.   The identity $\{\Psi_{F}, \Psi_{H} \} =0$
means that $\Psi_{H}$ belongs to the kernel of the background shift generated by $\{\Psi_{F}, *\}$. 
It can be understood that such string fields 
are not limited to $\Psi_H$ but fill large part of the space of string fields. 
A crucial observation is that both $\Psi_F$ and $\Psi_H$ are projected string fields: 
 \begin{equation}
  \Psi_{F}  = p_1 \Psi_{F} q_2, \qquad \Psi_{H}  = p_2 \Psi_{H} q_1,
 \end{equation}
where $p_{i}$ and $q_{i}$ $(i,j = 1,2)$ are star algebra projectors defined by
\begin{alignat}{2}
p_1 & = F c B H, & \quad q_{1} & = F B c H, \\
p_2 & = H c B F, & \quad q_{2} & = H B c F.
\end{alignat}
$p_{i}$ and $q_{i}$ are orthogonal projectors since $p_{i} + q_{i} =1$.  
Projectors with different indexes do not always 
commute but obey rather non-trivial
algebra which is summarized as
\begin{alignat}{4}
p_{i} p_{j}  & = p_{i}, &  \quad q_{i} q_{j} & = q_{j},\\
p_{i} q_{j}  & = 0,     &  \quad q_{i} p_{j} & = \epsilon_{i j} (p_{j} -p_{i}),
\end{alignat}  
where $\epsilon_{i j}$ is the antisymmetric tensor. 
Using these projectors, 
one can decompose arbitrary string field into projected sectors\footnote{%
	We are inspired by a work~\cite{Kishimoto:2014yea} in 
	which the string field is decomposed by the KMTT
	projectors.   
	}. 
Here we are interested in two kinds of decompositions.  One is ``D-like'' 
decomposition which is given by
\begin{equation}
\begin{split}
\Psi & =  (p_1+q_1) \Psi (p_2+ q_2) \\
& = p_1 \Psi p_2 + p_1 \Psi q_2 + 
q_1 \Psi p_2 + q_1 \Psi p_{2} \\
& = \psi_{1} + \psi_{2} + \psi_{3} + \psi_{4},
\end{split}
\end{equation}
and the other is ``gD-like'' decomposition
\begin{equation}
\begin{split}
\Psi & =  (p_2+q_2) \Psi (p_1+ q_1) \\
& = p_2 \Psi p_1 + p_2 \Psi q_1 + 
q_2 \Psi p_1 + q_2 \Psi p_{1} \\
& = \phi_{1} + \phi_{2} + \phi_{3} + \phi_{4}.
\end{split}
\label{gdlike}
\end{equation}
One can readily find that $\Psi_{D}= - \Psi_{F} $ and $\Psi_{gD} = \Psi_{H}$ are components of  
 $\psi_{2}$ and $\phi_{2}$ respectively.  Then, the orthogonality 
 between $\Psi_{F}$ and $\Psi_{H}$ is not 
 a limited to these solutions but 
 can be extended to the projected 
 sectors:
\begin{equation}
  \psi_{2} \phi_{2} = \phi_{2}\psi_{2} =0.
\end{equation}
Usually, one may implicitly assume that an nontrivial solution of equation of motion 
causes background shift for arbitrary string field. 
However, it is not true if that solution has a kernel.  
For illustration, we consider OSFT
action around the tachyon vacuum with expanding
the fluctuation in gD-like expansion \eqref{gdlike}: 
\begin{equation}
\Psi = \Psi_{F}  +  \phi_2  + \eta,
\end{equation}
where $\eta =  \phi_1 +\phi_3 + \phi_4$.
In this case, the string field $\phi_2$ is the 
kernel of the solution $\Psi_F$. The OSFT action is expanded as
\begin{equation}
 S [\Psi]  = \frac{1}{2} \Tr[\eta Q_{F} \eta] +
 \frac{1}{2} \Tr[\phi_2 Q_{B} \phi_2] + 
 \Tr [ \phi_2 Q_{B} \eta    ] +(\text{cubic}) + (\text{const.}).
 \label{paction}
\end{equation}
It is observed that kinetic terms with $\phi_2$
are not shifted to that for the tachyon vacuum ($Q_F$) but
remain un-shifted ($Q_B$).  
Therefore, $\phi_2$
can be interpreted as a degrees of freedom on a
``residual'' D-brane.  Equations of motion for the action \eqref{paction} are
\begin{align}
Q_B \phi_{2} + \phi_{2}^{2}+ 
Q_{B} \eta + \phi_{2} \eta + \eta^2 & = 0, \\
Q_F \eta + \eta^2 + 
Q_{B} \phi_2 + \phi_{2}^{2}  + \phi_2 \eta & = 0.
\end{align}
Since $\phi_2$ and $\eta$ are projected components, one can project out either
of them.  By setting $\eta=0$, both of the above equations reduce to 
the equation of motion for a D-brane
\begin{equation}
Q_B \phi_2 + \phi_{2}^{2} =0.
\end{equation}
Thus tachyon condensation takes place again with a solution
solution $\phi_2  =  \Psi_{H}$. 
The negative tension 
of the solution can be interpreted as a result of tachyon
condensation in the residual sector.  Thus existence of the residual sector
is responsible to the peculiar nature of gD-brane which has negative tension.  
One can argue whether the residual sector can be removed by gauge fixing.  
A condition that picks up $\eta$ component from \eqref{gdlike} reads
\begin{equation}
 p_{2} \Psi q_{1} = 0. \label{gaugefix}
\end{equation} 
This condition looks like a sort of linear $b$ gauge fixing~\cite{Kiermaier:2007jg}
since $p_{2}$ and $q_{1}$ include products of $B$ and $c$.  
In order to check validity of the condition, one has to show that arbitrary 
string field can be gauge transformed to satisfy \eqref{gaugefix}, 
and also that no residual gauge symmetry is left.   
Proof seems to require detailed and careful evaluation of gauge transformation, so
we leave it as a future task.  

\section{Multiple branes} \label{sec:multiple}

We next turn to constriction of a solution with arbitrary number of D and gD branes. 
Basic ingredients of our construction is the modified
boundary condition changing (BCC) operators~\cite{Erler:2014eqa}
\begin{equation}
    \Sigma_{a} = Q_{F} \left(F B \sigma_a F
    \right), \quad
     \overline{\Sigma}_{a} = Q_{F} \left(F B 
     \bar{\sigma}_a F \right),
\end{equation}
where $\sigma_a$ and $\bar{\sigma}$ are BCC operators 
associated with certain boundary conditions, and 
$a$ corresponds to Chan-Paton factor which labels D or gD brane.   By construction, they are  $Q_F$ exact therefore vanish when multiplied with it:
\begin{equation}
    Q_{F} \Sigma_a = Q_{F} \overline{\Sigma}_{a} =0.
    \label{Bcc1}
\end{equation}
They also inherit the algebra of original BCC operators: 
\begin{equation}
    \overline{\Sigma}_{a} \Sigma_{b} 
    = \delta_{a b},  \quad \overline{\Sigma}_{a} \Sigma_{b} = 
    \textbf{finite} \times \delta_{ab} 
    \label{Bcc2}
\end{equation}
We consider the theory at the tachyon vacuum whose kinetic operator is $Q_F$.  Then, it is easily understood that, if $\Phi$ is a solution of the equation of motion, $\Sigma_{a} \Phi \overline{\Sigma}_{a}$ is also a solution.  Therefore,  for given set of solutions $\Phi_1, \Phi_2 , \ldots, \Phi_{N}$,  we 
can  construct a set of mutually orthogonal solutions 
\begin{equation}
 \Sigma_{1} \Phi_1 \overline{\Sigma}_{1}, \ \Sigma_{2} \Phi_2 \overline{\Sigma}_{2},\ \ldots \ ,\Sigma_{N} \Phi_N \overline{\Sigma}_{N}.
\end{equation}
Of course, their sum is also a solution due to the orthogonality of the modified BCC projectors.  
The sum is conveniently described by vector notation:
\begin{equation}
 (\Phi_1, \Phi_2, \ldots, \Phi_n) = \sum_{a=1}^{N}  \Sigma_{a} \Phi_a  \overline{\Sigma}_{a} 
\end{equation}
We can easily construct solutions for multiple D 
or gD branes respectively by
\begin{equation}
 \Psi_{D}^{(N)} =  \underbrace{(\Psi_{D}, \Psi_{D}, \ldots, \Psi_{D})}_{N},  \label{nd}
\end{equation}
\begin{equation}
 \Psi_{gD}^{(N)} = \underbrace{(\Psi_{gD}, \Psi_{gD}, \ldots, \Psi_{gD})}_{N}, \label{ngd}
\end{equation}
where $\Psi_{D}$ and $\Psi_{gD}$ are brane solutions 
 defined by \eqref{Dsolution} and
\eqref{gDsolution} respectively. 
Due to the orthogonality, gauge invariant observables split into pieces for each component.  
For example, classical action for D-branes is evaluated as
\begin{align}
 S_{F} [\Psi_{D}^{(N)}] & =  N S_{F} [\Psi_{D}] \\
  & = N. 
\end{align}
Similar evaluation for gD-branes derives the value $-N$.   
Then the cancellation between tensions can be confirmed as follows.  
We 
First we construct a pair of $N$ D-branes and $N$ gD-branes:
note that $\Psi_{D}^{(N)} +  \Psi_{gD}^{(N)} $ again becomes a solution since
\begin{equation}
\begin{split}
 \Psi^{(N)}_{D+gD} & =\Psi_{D}^{(N)} +  \Psi_{gD}^{(N)} \\
 & = \underbrace{
 (\Psi_{D+gD}, \Psi_{D+gD}, \ldots, \Psi_{D+gD})}_{N} \\ 
\end{split}
\end{equation}
where $\Psi_{D+gD} = \Psi_{D} + \Psi_{gD}$  is the solution 
for a D-gD pair defined by \eqref{D+gDsolution}.
Then, the tension cancels as
\begin{equation}
\begin{split}
S_{F} [\Psi_{D+gD}^{(N)} ] & =  S_{F}[\Psi_{D}^{(N)} ] + S_{F}[\Psi_{gD}^{(N)} ]\\
& = + N -  N \\
& = 0.
\end{split}
\end{equation}
Similar cancellations also hold for other observables.    

We next ask whether the cancellation of cohomology occurs between D and gD branes.   To see this, let us consider the kinetic operator around  $\Psi^{(N)}_{D+gD}$. 
  It is given by
\begin{equation}
 Q^{(N)}_{D+gD} \Psi = Q_{F} \Psi + \{\Psi_{D+gD}^{(N)},\Psi\}. \label{nDgDcharge}
\end{equation}
Right hand side of \eqref{nDgDcharge} does not seem to
correspond to any known operator with empty cohomology.  
However, it can be shown that this operator splits into $n$ copies of kinetic operator with empty cohomology for particular
choice of $\Psi$.  Such choice of a string field is namely given by a vector of same type as \eqref{nd} or \eqref{ngd}:
\begin{equation}
 \Psi^{(N)} = \underbrace{(\Psi_1, \Psi_2, \ldots, \Psi_{N})}_{N}. \label{psiprojected}
\end{equation}
Note that this choice corresponds to a subset of projected string fields considered in \cite{Kishimoto:2014yea}. 
Then, with \eqref{Bcc1} and \eqref{Bcc2}, it is straightforward to show that
\begin{equation}
 Q_{D+gD}^{(N)} \Psi^{(N)} =  
 \underbrace{(Q_{D+gD} \Psi_1, Q_{D+gD} \Psi_2, \ldots, Q_{D+gD} \Psi_N)}_{N},
\end{equation}
where $Q_{D+gD}$ is a kinetic operator for a D-gD pair
given by \eqref{DgDcharge}.  
Thus the kinetic operator around $N$ D-gD pair splits into $Q_{D+gD}$ which has no cohomology. 
Therefore, cancellation between cohomology holds for the 
projected string field \eqref{psiprojected}. 

A solution with different numbers of D and gD branes can be constructed similarly. For example, consider
\begin{align}
 \Psi^{(N+K)}_{D} & = (\underbrace{\Psi_D, \Psi_D, \ldots, \Psi_{D}}_{N}, \underbrace{\Psi_D, \Psi_D, \ldots, \Psi_{D}}_{K},
   \underbrace{0,0,\ldots,0}_{M} ),\\
 \Psi^{(K+M)}_{gD} & = (\underbrace{0, 0, \ldots, 0}_{N}, \underbrace{\Psi_{gD}, \Psi_{gD}, \ldots, \Psi_{gD}}_{K},
   \underbrace{\Psi_{gD}, \Psi_{gD}, \ldots, \Psi_{gD}}_{M}).
\end{align}
A sum $\Psi^{(N+K)}_{D} +  \Psi^{(K+M)}_{gD} $ is a
solution that represents $N+K$ D-branes and $K+M$ gD-branes.   
Since $\Psi_{D}+ \Psi_{gD} = \Psi_{D+gD}$, the sum is written as
\begin{equation}
\begin{split}
\Psi^{(N,K,M)} & =
 \Psi^{(N+K)}_{D} +  \Psi^{(K+M)}_{gD} \\
 & = (\underbrace{\Psi_D, \Psi_D, \ldots, \Psi_{D}}_{N}, \underbrace{\Psi_{D+gD}, \Psi_{D+gD}, \ldots, \Psi_{D+gD}}_{K},
   \underbrace{\Psi_{gD},\Psi_{gD} ,\ldots, \Psi_{gD}}_{M} ). 
   \end{split}
   \label{multiDgD}
\end{equation}
The value of the classical action is given by $N-M$ as expected.  However, the cohomology
cancel only in $k$ slots in the middle of the vector \eqref{multiDgD}.   This can be
shown as follows. We introduce a shorthand notation
\begin{equation}
\Phi = (\Phi^{(N)}, \Phi^{(K)}, \Phi^{(M)} )
\end{equation}
where three components stand for vectors in each $N, K$ and $M$ slots.  
Then, kinetic operator around $\Psi^{(N,K,M)}$ is evaluated as
\begin{equation}
Q_{F} \Phi + \{\Psi^{(N,K,M)}, \Phi\} = 
(Q_{B} \Phi^{(N)}, Q_{D+gD} \Phi^{(K)},  Q_{gD}  \Phi^{(M)} ),
\end{equation}
where $Q_{gD} = Q_{F}+ \{\Psi_{gD},* \}$ is a kinetic operator for
a gD-brane. 
It is obvious that only $Q_{D+gD}$ has vanishing cohomology therefore
cancellation occurs only in the middle $K$ slots.  First $N$ and last $M$ slots 
have no chance to cancel since their Chan-Paton factors do not overlap.    
  
We can further extend above system by introducing off-diagonal Chan-Paton factors
of $\Phi$ following with the method of~\cite{Erler:2014eqa,Kishimoto:2014yea}.  
We introduce KMTT projectors
$P_{k}=\Sigma_k \overline{\Sigma}_k$ and arrange them as
\begin{equation}
  \{P_{0}, P_{a},  P_{\alpha},   P_{A} \},
\end{equation}
where $a$, $\alpha$, $A$ are assigned to three slots of indexes in 
\eqref{multiDgD},  and $P_{0} =  1- \sum_{k=1}^{N+K+M}\Sigma_k \overline{\Sigma}_k$
is a complementary projector.  These labels are identified 
with Chan-Paton factors for each branes; $0$, $a$, $\alpha$, $A$ are assigned
to a tachyon vacuum, $N$ D-branes, $K$ tachyon vacuua and $M$ gD-branes respectively.  
This assignment comes from the fact that a component of string field in each sector obeys
appropriate kinetic term in the OSFT action expanded around  \eqref{multiDgD}. 
For example, as for $\Phi_{ab} = P_{a} \Psi P_{b} $,
 \begin{equation}
 Q_{F} \Phi_{ab} + \{ \Psi^{(N,K,M)}, \Phi_{ab}   \}
 =  Q_{B} \Phi_{ab}
 \end{equation}
holds.  The component $\Phi_{ab}$ corresponds to open string field between D-brane
$a$ and $b$ since $Q_B$ is a kinetic operator for a D-brane. 
A component with mixed indexes, like $\Phi_{a A}$, is interpreted as
a string field which connects different kinds of branes.   

\section{Conclusions} \label{sec:conclusions}
In this paper, we have studied ghost D-branes in the context of open string field theory. 
First, we constructed the classical solutions for the D-gD system
with a help of the MNT solutions.   We have confirmed cancellation of gauge invariant
observables for the D-gD pair.  It have been shown that the BRST cohomology of a
D-gD pair cancels.   Next we have extended previous result to a system 
with arbitrary numbers of D and gD branes. We have constructed corresponding solution
using modified BCC operators~\cite{Erler:2014eqa}.   
It have shown that cancellation of cohomology holds for branes with 
common Chan-Paton factors.  

We would like to answer the two questions asked in introduction. 
Fist question asks whether gD-branes are physical objects. 
While we did not find conclusive answer to this, 
we found that gD-branes belong to a sector of open string field which is not affected by 
the shift of background.  As mentioned in section \ref{sec:MNT}, we can ask whether 
such sector is gauge away.  If so, gD-branes turn
out to be unphysical objects and we do not need to worry about negative tension.
On the other hand, severe problem will remain if the gD-sector cannot be removed.
Multiple gD-branes constructed in \ref{sec:multiple} 
make the spectrum of OSFT unbounded below.   
Therefore ,it remains important to give conclusive answer to this question.

Second question asks difference between the original picture~\cite{Okuda:2006fb}
of Okuda Takayanagi and OSFT prescription.  
We have confirmed that the BRST cohomology cancels between D and gD branes. 
Therefore quantum fluctuations around a D-gD pair cancel expectedly. 
However, unlike the original picture, the cancellation 
cannot be extend to the whole partition function.   
This discrepancy leads rather peculiar phenomenon in OSFT.
Let us consider multiple D-gD system.  The kinetic
operator of a D-gD pair is given by $O_{D+gD}=Q_{H}$ which is equivalent
to that of tachyon vacuum.  One can consider an effective action
for remaining branes by integrating out string fields for
canceled D-gD pairs.  
As has been conjectured earlier~\cite{Gaiotto:2001ji,Drukker:2002ct,
	Drukker:2003hh,Takahashi:2003kq}, 
such integration will leave closed string amplitudes with no boundaries, i.e., 
closed string tadpoles.  
Therefore, infinitely many D-gD pair implies 
infinitely many tadpoles. Although this seems rather pathological, useful
applications and implications will be expected.

Finally, we present rather positive perspective of our result.  
Since a D-gD pair is equivalent to the tachyon (or the closed string) vacuum, we 
can say that the closed string vacuum is described by 
two vacua in different BCFTs, i.e., one for D-brabe and the other for gD-brane.
It is expected that an extension to excited states of closed string will
leads brand-new formulation of closed string theory in terms of open string. 
Even if gD-brane is unphysical, it will play a role of auxiliary degrees 
of freedom which is useful for such formulation.  
  
To push this program forward, we should identify BCFT for gD-brane.
Naively, it is expected that a gD-brane carries same boundary condition as
that of a D-brane since boundary states only differ in their sign.  In order to 
identify gD-brane BCFT in OSFT context, one has to derive cohomology
of gD-brane.
However, direct identification of it seems to be not straightforward, as 
some attempts presented below indicate. 
For example, formal homotopy operator of the gD kinetic operator
	\begin{equation}
	Q_{F+H} = Q_{B} + \{\Psi_{F} + \Psi_{H}, *    \}.
	\end{equation}
seems to vanish at least within $KBc$ algebra. This is not 
similar to the D-brane cohomology whose $B/K$ is a formal homotopy
operator.  Another but related attempt is finding left and right 
transformations~\cite{Erler:2012qn,Erler:2012qr}
which connects D and gD-branes.  This had already derived by
MNT~\cite{Masuda:2012kt} as
	\begin{align*}
	U_{L}  & = M K (\frac{F}{1-F^2}) c B (\frac{1-F^2}{F} ) \\
	& =  M K J c B J^{-1}.
	\end{align*}  
	where $J = F/(1-F^2)$.   The latter piece $JcBJ^{-1}$ is a star algebra projector
	therefore has a nontrivial kernel regardless of potential singularity due to
	$K=0$ poles of $M$ and $F^{-1}$. Cohomologies of 
	D and gD branes may be related in projected space of string fields obtained
	by excluding this 
	kernel.  To validate this discussion, it is necessary to confirm
	whether the space obtained by projecting out the kernel is suitable
	to describe D or gD brane BCFT.   
  
\section*{Acknowledgements}

The author world like to thank organizers of the conference 
``Progress in Quantum Field Theory and String Theory II''
at Osaka city university for an opportunity 
of poster presentation. Discussions with participants 
of the conference were quite useful to complete our research.
In particular, we would like to thank Masako Asano, 
Hiroshi Itoyama, Isao Kishimoto and Keisuke Ohashi for valuable discussions.  
A support from JSPS/RFBR bilateral project ``faces of matrix models in quantum field theory
and statistical mechanics'' is gratefully appreciated.

\section*{Conflict of Interest Statement}

The author declares that there is no conflict of interest regarding the publication of this paper.

\bibliographystyle{utphys}
\bibliography{zeze}

\providecommand{\href}[2]{#2}\begingroup\raggedright\begin{thebibliography}{10}

\bibitem{Witten:1985cc}
E.~Witten, ``{Noncommutative Geometry and String Field Theory},''
\href{http://dx.doi.org/10.1016/0550-3213(86)90155-0}{{\em Nucl. Phys.}
  {\bfseries B268} (1986) 253--294}.

\bibitem{Erler:2014eqa}
T.~Erler and C.~Maccaferri, ``{String Field Theory Solution for Any Open String
  Background},'' \href{http://dx.doi.org/10.1007/JHEP10(2014)029}{{\em JHEP}
  {\bfseries 10} (2014) 029},
\href{http://arxiv.org/abs/1406.3021}{{\ttfamily arXiv:1406.3021 [hep-th]}}.

\bibitem{Cardy:2004hm}
J.~L. Cardy, ``{Boundary conformal field theory},''
\href{http://arxiv.org/abs/hep-th/0411189}{{\ttfamily arXiv:hep-th/0411189
  [hep-th]}}.

\bibitem{Kojita:2016jwe}
T.~Kojita, C.~Maccaferri, T.~Masuda, and M.~Schnabl, ``{Topological defects in
  open string field theory},''
\href{http://arxiv.org/abs/1612.01997}{{\ttfamily arXiv:1612.01997 [hep-th]}}.

\bibitem{Okuda:2006fb}
T.~Okuda and T.~Takayanagi, ``{Ghost D-branes},''
  \href{http://dx.doi.org/10.1088/1126-6708/2006/03/062}{{\em JHEP} {\bfseries
  03} (2006) 062},
\href{http://arxiv.org/abs/hep-th/0601024}{{\ttfamily arXiv:hep-th/0601024
  [hep-th]}}.

\bibitem{Terashima:2006qm}
S.~Terashima, ``{Ghost D-brane, supersymmetry and matrix model},''
  \href{http://dx.doi.org/10.1088/1126-6708/2006/05/067}{{\em JHEP} {\bfseries
  05} (2006) 067},
\href{http://arxiv.org/abs/hep-th/0602271}{{\ttfamily arXiv:hep-th/0602271
  [hep-th]}}.

\bibitem{Evans:2006eq}
N.~Evans, T.~R. Morris, and O.~J. Rosten, ``{Gauge invariant regularization in
  the AdS/CFT correspondence and ghost D-branes},''
  \href{http://dx.doi.org/10.1016/j.physletb.2006.02.055}{{\em Phys. Lett.}
  {\bfseries B635} (2006) 148--150},
\href{http://arxiv.org/abs/hep-th/0601114}{{\ttfamily arXiv:hep-th/0601114
  [hep-th]}}.

\bibitem{Vafa:2001qf}
C.~Vafa, ``{Brane / anti-brane systems and U(N|M) supergroup},''
\href{http://arxiv.org/abs/hep-th/0101218}{{\ttfamily arXiv:hep-th/0101218
  [hep-th]}}.

\bibitem{Parkhomenko:2001ki}
S.~E. Parkhomenko, ``{BRST construction of D-branes in SU(2) WZW model},''
  \href{http://dx.doi.org/10.1016/S0550-3213(01)00444-8}{{\em Nucl. Phys.}
  {\bfseries B617} (2001) 198--214},
\href{http://arxiv.org/abs/hep-th/0103142}{{\ttfamily arXiv:hep-th/0103142
  [hep-th]}}.

\bibitem{Parkhomenko:2003gy}
S.~E. Parkhomenko, ``{Free field construction of D branes in N=2 superconformal
  minimal models},''
  \href{http://dx.doi.org/10.1016/j.nuclphysb.2003.08.032}{{\em Nucl. Phys.}
  {\bfseries B671} (2003) 325--342},
\href{http://arxiv.org/abs/hep-th/0301070}{{\ttfamily arXiv:hep-th/0301070
  [hep-th]}}.

\bibitem{Parkhomenko:2004ab}
S.~E. Parkhomenko, ``{Free field approach to D-branes in Gepner models},''
  \href{http://dx.doi.org/10.1016/j.nuclphysb.2005.10.011}{{\em Nucl. Phys.}
  {\bfseries B731} (2005) 360--388},
\href{http://arxiv.org/abs/hep-th/0412296}{{\ttfamily arXiv:hep-th/0412296
  [hep-th]}}.

\bibitem{Parkhomenko:2008dt}
S.~E. Parkhomenko, ``{Free Field Construction of D-Branes in Rational Models of
  CFT and Gepner Models},''
  \href{http://dx.doi.org/10.3842/SIGMA.2008.025}{{\em SIGMA} {\bfseries 4}
  (2008) 025},
\href{http://arxiv.org/abs/0802.3445}{{\ttfamily arXiv:0802.3445 [hep-th]}}.

\bibitem{Dijkgraaf:2016lym}
R.~Dijkgraaf, B.~Heidenreich, P.~Jefferson, and C.~Vafa, ``{Negative Branes,
  Supergroups and the Signature of Spacetime},''
\href{http://arxiv.org/abs/1603.05665}{{\ttfamily arXiv:1603.05665 [hep-th]}}.

\bibitem{Murata:2011ep}
M.~Murata and M.~Schnabl, ``{Multibrane Solutions in Open String Field
  Theory},'' \href{http://dx.doi.org/10.1007/JHEP07(2012)063}{{\em JHEP}
  {\bfseries 07} (2012) 063},
\href{http://arxiv.org/abs/1112.0591}{{\ttfamily arXiv:1112.0591 [hep-th]}}.

\bibitem{Hata:2011ke}
H.~Hata and T.~Kojita, ``{Winding Number in String Field Theory},''
  \href{http://dx.doi.org/10.1007/JHEP01(2012)088}{{\em JHEP} {\bfseries 01}
  (2012) 088},
\href{http://arxiv.org/abs/1111.2389}{{\ttfamily arXiv:1111.2389 [hep-th]}}.

\bibitem{Hata:2012cy}
H.~Hata and T.~Kojita, ``{Singularities in K-space and Multi-brane Solutions in
  Cubic String Field Theory},''
  \href{http://dx.doi.org/10.1007/JHEP02(2013)065}{{\em JHEP} {\bfseries 02}
  (2013) 065},
\href{http://arxiv.org/abs/1209.4406}{{\ttfamily arXiv:1209.4406 [hep-th]}}.

\bibitem{Masuda:2012kt}
T.~Masuda, T.~Noumi, and D.~Takahashi, ``{Constraints on a class of classical
  solutions in open string field theory},''
  \href{http://dx.doi.org/10.1007/JHEP10(2012)113}{{\em JHEP} {\bfseries 10}
  (2012) 113},
\href{http://arxiv.org/abs/1207.6220}{{\ttfamily arXiv:1207.6220 [hep-th]}}.

\bibitem{Erler:2012dz}
T.~Erler, ``{The Identity String Field and the Sliver Frame Level Expansion},''
  \href{http://dx.doi.org/10.1007/JHEP11(2012)150}{{\em JHEP} {\bfseries 11}
  (2012) 150},
\href{http://arxiv.org/abs/1208.6287}{{\ttfamily arXiv:1208.6287 [hep-th]}}.

\bibitem{Erler:2012qn}
T.~Erler and C.~Maccaferri, ``{Connecting Solutions in Open String Field Theory
  with Singular Gauge Transformations},''
  \href{http://dx.doi.org/10.1007/JHEP04(2012)107}{{\em JHEP} {\bfseries 04}
  (2012) 107},
\href{http://arxiv.org/abs/1201.5119}{{\ttfamily arXiv:1201.5119 [hep-th]}}.

\bibitem{Baba:2006rs}
Y.~Baba, N.~Ishibashi, and K.~Murakami, ``{D-branes and closed string field
  theory},'' \href{http://dx.doi.org/10.1088/1126-6708/2006/05/029}{{\em JHEP}
  {\bfseries 05} (2006) 029},
\href{http://arxiv.org/abs/hep-th/0603152}{{\ttfamily arXiv:hep-th/0603152
  [hep-th]}}.

\bibitem{Baba:2007tc}
Y.~Baba, N.~Ishibashi, and K.~Murakami, ``{D-brane States and Disk Amplitudes
  in OSp Invariant Closed String Field Theory},''
  \href{http://dx.doi.org/10.1088/1126-6708/2007/10/008}{{\em JHEP} {\bfseries
  10} (2007) 008},
\href{http://arxiv.org/abs/0706.1635}{{\ttfamily arXiv:0706.1635 [hep-th]}}.

\bibitem{Okawa:2006vm}
Y.~Okawa, ``{Comments on Schnabl's analytic solution for tachyon condensation
  in Witten's open string field theory},''
  \href{http://dx.doi.org/10.1088/1126-6708/2006/04/055}{{\em JHEP} {\bfseries
  04} (2006) 055},
\href{http://arxiv.org/abs/hep-th/0603159}{{\ttfamily arXiv:hep-th/0603159
  [hep-th]}}.

\bibitem{Kiermaier:2008qu}
M.~Kiermaier, Y.~Okawa, and B.~Zwiebach, ``{The boundary state from open string
  fields},''
\href{http://arxiv.org/abs/0810.1737}{{\ttfamily arXiv:0810.1737 [hep-th]}}.

\bibitem{Ellwood:2008jh}
I.~Ellwood, ``{The Closed string tadpole in open string field theory},''
  \href{http://dx.doi.org/10.1088/1126-6708/2008/08/063}{{\em JHEP} {\bfseries
  08} (2008) 063},
\href{http://arxiv.org/abs/0804.1131}{{\ttfamily arXiv:0804.1131 [hep-th]}}.

\bibitem{Kawano:2008ry}
T.~Kawano, I.~Kishimoto, and T.~Takahashi, ``{Gauge Invariant Overlaps for
  Classical Solutions in Open String Field Theory},''
  \href{http://dx.doi.org/10.1016/j.nuclphysb.2008.05.025}{{\em Nucl. Phys.}
  {\bfseries B803} (2008) 135--165},
\href{http://arxiv.org/abs/0804.1541}{{\ttfamily arXiv:0804.1541:0804.1541
  [hep-th]}}.

\bibitem{Kudrna:2012re}
M.~Kudrna, C.~Maccaferri, and M.~Schnabl, ``{Boundary State from Ellwood
  Invariants},'' \href{http://dx.doi.org/10.1007/JHEP07(2013)033}{{\em JHEP}
  {\bfseries 07} (2013) 033},
\href{http://arxiv.org/abs/1207.4785}{{\ttfamily arXiv:1207.4785 [hep-th]}}.

\bibitem{Ellwood:2006ba}
I.~Ellwood and M.~Schnabl, ``{Proof of vanishing cohomology at the tachyon
  vacuum},'' \href{http://dx.doi.org/10.1088/1126-6708/2007/02/096}{{\em JHEP}
  {\bfseries 02} (2007) 096},
\href{http://arxiv.org/abs/hep-th/0606142}{{\ttfamily arXiv:hep-th/0606142
  [hep-th]}}.

\bibitem{Kishimoto:2014yea}
I.~Kishimoto, T.~Masuda, T.~Takahashi, and S.~Takemoto, ``{Open String Fields
  as Matrices},'' \href{http://dx.doi.org/10.1093/ptep/ptv023}{{\em PTEP}
  {\bfseries 2015} no.~3, (2015) 033B05},
\href{http://arxiv.org/abs/1412.4855}{{\ttfamily arXiv:1412.4855 [hep-th]}}.

\bibitem{Kiermaier:2007jg}
M.~Kiermaier, A.~Sen, and B.~Zwiebach, ``{Linear b-Gauges for Open String
  Fields},'' \href{http://dx.doi.org/10.1088/1126-6708/2008/03/050}{{\em JHEP}
  {\bfseries 03} (2008) 050},
\href{http://arxiv.org/abs/0712.0627}{{\ttfamily arXiv:0712.0627 [hep-th]}}.

\bibitem{Gaiotto:2001ji}
D.~Gaiotto, L.~Rastelli, A.~Sen, and B.~Zwiebach, ``{Ghost structure and closed
  strings in vacuum string field theory},'' {\em Adv. Theor. Math. Phys.}
  {\bfseries 6} (2003) 403--456,
\href{http://arxiv.org/abs/hep-th/0111129}{{\ttfamily arXiv:hep-th/0111129
  [hep-th]}}.

\bibitem{Drukker:2002ct}
N.~Drukker, ``{Closed string amplitudes from gauge fixed string field
  theory},'' \href{http://dx.doi.org/10.1103/PhysRevD.67.126004}{{\em Phys.
  Rev.} {\bfseries D67} (2003) 126004},
\href{http://arxiv.org/abs/hep-th/0207266}{{\ttfamily arXiv:hep-th/0207266
  [hep-th]}}.

\bibitem{Drukker:2003hh}
N.~Drukker, ``{On different actions for the vacuum of bosonic string field
  theory},'' \href{http://dx.doi.org/10.1088/1126-6708/2003/08/017}{{\em JHEP}
  {\bfseries 08} (2003) 017},
\href{http://arxiv.org/abs/hep-th/0301079}{{\ttfamily arXiv:hep-th/0301079
  [hep-th]}}.

\bibitem{Takahashi:2003kq}
T.~Takahashi and S.~Zeze, ``{Closed string amplitudes in open string field
  theory},'' \href{http://dx.doi.org/10.1088/1126-6708/2003/08/020}{{\em JHEP}
  {\bfseries 08} (2003) 020},
\href{http://arxiv.org/abs/hep-th/0307173}{{\ttfamily arXiv:hep-th/0307173
  [hep-th]}}.

\bibitem{Erler:2012qr}
T.~Erler and C.~Maccaferri, ``{The Phantom Term in Open String Field Theory},''
  \href{http://dx.doi.org/10.1007/JHEP06(2012)084}{{\em JHEP} {\bfseries 06}
  (2012) 084},
\href{http://arxiv.org/abs/1201.5122}{{\ttfamily arXiv:1201.5122 [hep-th]}}.

\end{thebibliography}\endgroup

\end{document}